\documentclass[prl,showpacs,twocolumn]{revtex4}

\usepackage{graphicx}

\begin{document}
\preprint{DRAFT}
\bibliographystyle{prsty}

\title{Plasmonic crystal demultiplexer and multiports}

\author{Aurelien Drezet}
\author{Daniel Koller}
\author{Andreas Hohenau}
\author{Alfred Leitner}
\author{Franz R. Aussenegg}
\author{Joachim R. Krenn}

\affiliation{Institute of Physics and Erwin Schr\"{o}dinger Institute for
Nanoscale Research, Karl--Franzens--University, Universit\"{a}tsplatz 5,
8010 Graz, Austria}

\date{\today}

\maketitle

\textbf{Artificially built periodic optical structures in
dielectric and metallic media have generated considerable interest
due to their potential for optical device miniaturization
\cite{PC,Smith,Barnes:2003,Ebbesen}. In this context plasmonics,
i.e., optics based on surface plasmon polaritons (SPPs)
\cite{Barnes:2003} offers new exciting prospects. SPPs are hybrid
light/electron surface waves at the interface between a dielectric
and a metal \cite{Raether:1988} and as such hold the potential for
2D optical functionality. Indeed, SPP elements as mirrors,
splitters and interferometers have been recently demonstrated
\cite{Ditlbacher:2002,Weeber:2005,Stepanov:2005,Drezet:2005,Drezet:2006}.
However, for plasmonics to qualify at the information technology
level requires necessarily the realization of wavelength division
(demultiplexing) which constitutes a fundamental ingredient of
optical communication. In the following we experimentally
demonstrate  2D SPP demultiplexing in the visible spectral range
by using photonic crystals for SPPs (plasmonic crystals). In
addition, we demonstrate that plasmonic crystal are capable of
realizing integrated linear multiports which could constitute
building blocks of analog or quantum optical computing.}

The introduction of 2D plasmonic crystals creates new
opportunities for nanooptics since such metal-based structures can
generate huge optical band gaps \cite{Weeber:2006}. Indeed,
complete SPP confinement (i.~e., an all-angle optical band gap)
has been demonstrated \cite{Barnes:1996} which opens up a venue
for SPP based waveguiding \cite{Bozhevolnyi:2001}. Simpler
geometries as 1D plasmonic band gap structures (Bragg mirrors or
gratings) are already well mastered, typical SPP Bragg reflectors
being built from periodically arranged lines of nanoscale metal
protrusions or indentations on a metal film
\cite{Ditlbacher:2002,Weeber:2005}. Bragg diffraction occurs when
the in--plane wave vector $\mathbf{k}_{0}$ of the incoming SPP
wave can pick up a momentum $\mathbf{G}=\pm p 2 \pi \mathbf{e}/d$.
$\mathbf{G}$ is a reciprocal vector of the Bragg lattice oriented
along the periodicity direction of the mirror, $d$ is the lattice
period and p is an integer. The diffracted wave vector must
consequently fulfill
\begin{equation}
\mathbf{k}_{d}=\mathbf{k}_{0} + \mathbf{G}.
\end{equation}
In addition, energy conservation imposes the condition
$|\mathbf{k}_{d}|=|\mathbf{k}_{0}|$ and as a consequence Bragg
reflection occurs only for specific angles and wavelengths. This
opens a possibility for spectral demultiplexing if we put
different Bragg mirrors 1, 2,... in series such that for each
mirror Eq.~1 is satisfied at a different SPP wavelengths
$\lambda_{1}$, $\lambda_{2}$,... . For practical reasons, however,
we do not use this configuration here due to the short SPP
propagation length $L_{SPP}<50 $ $\mu$m in the visible spectral
range. We propose rather a compact solution in which two Bragg
mirrors optimized for two different wavelengths are superimposed.
Indeed, if nanoscale protrusions (or indentations) are positioned
at the intersection points of two sets of Bragg lines with
different orientations built to reflect respectively the
wavelengths $\lambda_{1}$ and $\lambda_{2}$ one obtains a 2D
plasmonic crystal presenting optical band gaps around these
wavelengths (Figs.~1a,b).

To demonstrate this approach we have investigated a plasmonic
crystal demultiplexer optimized for the two SPP wavelengths
$\lambda_{1}\simeq 784$ nm and $\lambda_{2}\simeq 730$ nm which
correspond to laser wavelengths of respectively 800 nm and 750 nm.
We fabricated rectangular 2D lattices made of gold protrusions
(200 nm diameter, 50 nm height) deposited on a gold film (see
Fig.~1a). As sketched in Fig.~1b we expect in such conditions that
incoming SPP waves with wavelength $\lambda_{1}$ and $\lambda_{2}$
are reflected in two opposite directions. This is indeed what is
observed experimentally (compare Fig.~1c and 1e) using leakage
radiation microscopy (LRM) which is based on the conversion of SPP
waves to propagating light
\cite{Hecht:1996,Bouhelier:2001,Stepanov:2005}.

To understand more precisely this effect we must apply Eq.~1
considering the reciprocal vector
$\mathbf{G}=p_{1}\mathbf{f}_{1}+p_{2}\mathbf{f}_{2}$. The
reciprocal lattice basis $\mathbf{f}_{i}$ (i=1,2) is connected to
the primitive lattice vector basis $\mathbf{e_{i}}$ (see Fig.~1a)
by $\mathbf{f}_{i}=-2\pi/d^{2}_{i}\mathbf{e}_{i}$, where $d_{i}$
are the lattice periods $|\mathbf{e}_{i}|=d_{i}$. Values of
$d_{1}=\lambda_{1}/\sqrt{2}= 554$ nm and
$d_{2}=\lambda_{2}/\sqrt{2}=516$ nm were chosen in order to
realize $45^{\circ}$ Bragg reflection. An incident SPP impinging
on the crystal in the vertical direction $[+1,+1]$ (see Fig.~1a)
is thus expected to be reflected to the right (i.~e., the
$[+1,-1]$ direction, see Fig.~1b) or to the left (i.~e., the
$[-1,+1]$ direction, see Fig.~1b) depending on the SPP wavelength
$\lambda_{2}$ or $\lambda_{1}$, respectively. SPPs are launched by
focussing a laser beam onto a gold ridge \cite{Ditlbacher:2002}
and propagate subsequently in the directions $[ \pm 1,\pm 1]$, as
seen in Figs.~1c,e which show the LRM images for the SPP
wavelengths $\lambda_{2}$ and $\lambda_{1}$. We observe that the
SPP propagating in the direction $[+1,+1]$ is efficiently
reflected upon interaction with the plasmonic crystal as expected,
i.e., to the right for $\lambda_{2}$ and to the left for
$\lambda_{1}$ (Figs.~1c,e).

Being a far-field optical method LRM enables the imaging and
direct quantitative analysis of SPPs in both direct and Fourier
space \cite{Hecht:1996,Drezet:2006}. The Fourier image (i.e.,
momentum maps) corresponding to Fig.~1c (respectively 1e) is shown
in Fig.~1d (respectively 1f) and reveals three distinct spots. The
spots labelled 1 and 3 correspond to the SPPs propagating in the
$[+1,+1]$ and $[-1,-1]$ directions, respectively, while the spot
labelled 2 corresponds to the reflected SPP (Figs.~1d). Due to
energy conservation all three spots 1-3 must be located on a same
Ewald circle, as found in the experiment. The transfer of momentum
(i.~e., $\mathbf{f}_{i}$ at the SPP wavelength $\lambda_{i}$) is
directly evident from the images as plotted by the dashed arrows
in Figs.~1d,f while incident and reflected SPPs are represented by
the continuous arrows.

Furthermore, these data allow the strightforward quantitative
analysis of the plasmonic crystal efficiency. For this purpose, we
have to take into account the exponential damping of the SPP
intensity along the SPP propagation direction. In quantitative
terms, the intensities of the three spots 1-3 in Figs.~1d,f are
related by the expressions $I_{2}=R e^{-r/L_{SPP}}I_{3}$ and
$I_{1}=(1-(1-T) e^{-r/L_{SPP}})I_{3}$, where $R$ and $T$ are the
reflection and transmission coefficients of the plasmonic crystal
and r=32$\mu m$ is the distance separating the SPP launch ridge
from the plasmonic crystal.  On the basis of cross-cuts trough the
spots 1-3 to retrieve quantitative information from the images
(see Fig.~1g) we deduced $R\simeq 80$\% and $T\simeq 5$\% for the
SPP wavelength $\lambda_{1}$ \cite{Drezet:2006}. This implies
losses $S=1-T-R$ of around 20\% which we attribute to out-of-plane
scattering. The radial shape of the observed SPP peaks (see
Fig.~1g) can be theoretically reproduced by using a Lorentzian
fit. This is in agreement with previous studies
\cite{Burke:1986,Drezet:2006} showing that the
full-width-at-half-maximum of such peaks are a direct measure of
$1/L_{SPP}$. We found here $L_{SPP}=30$ $\mu m$ which is in good
agreement with the measurement of the exponential SPP damping in
direct space (not shown). Angular cross--cuts of the SPP peaks
(see Fig.~1h) show that the angular divergence is conserved when
going from 1 to 3. The same analysis for the SPP wavelength
$\lambda_{2}$ yields $R\simeq 70$\%, $T\simeq 10$\% and $S\simeq
30$\%.

To corroborate our results we simulated the interaction of a SPP
beam with the plasmonic crystal by using a 2D dipolar model
\cite{Drezet:2005}. While interparticle coupling was neglected in
our model (i.~e.~we worked in the limit of the first Born
approximation) qualitative agreement with the experimental
observations for the two considered wavelengths is obtained, see
the insets in Figs.~1c,e.

We now turn to the special case $\lambda_{1}=\lambda_{2}$ (and thus a
plasmonic crystal with a quadratic unit cell) which corresponds to a SPP
splitter (Figs.~2a,b). We studied this system for a laser wavelength of 750 nm
corresponding to a SPP wavelength $\lambda_{1}=\lambda_{2}=730$ nm. In the
direct space LRM image in Fig.~2c the splitting of the incoming SPP beam is
clearly observed. In the Fourier image in Fig.~2d one observes 4 spots
corresponding to the two counter--propagating SPPs launched from the gold
ridge and to the two reflected SPP beams. This image again illustrates
directly how momentum transfer operates in such a device. Quantitative
analysis as above indicates that about 80 \% of the incoming SPP intensity is
reflected equally between SPPs propagating to the right and to the left of
the crystal (i.~e., in the $[+1,-1]$ and $[-1,+1]$ directions,
respectively). The residual part of the intensity ($S\simeq 20$ \%) is again
attributed to loss due to scattering.

The beam splitting and demultiplexing properties of plasmonic crystals are not
restricted to two output beams. For example, choosing a triangular lattice
instead of a rectangular one (Fig.~3a) splits an input beam into three output
beams. The triangular lattice can be thought of as resulting from the overlap
of three Bragg mirrors such that the angle between the corresponding Bragg
lines equals $60^{\circ}$. Choosing the distance $d$ separating two adjacent
parallel Bragg lines such that $d=\lambda_{SPP}/\sqrt{3}$ one obtains Bragg
reflection at $30^{\circ}$ incidence angle. This corresponds to an
interparticle distance in the triangular lattice of $a=2\lambda_{SPP}/3$. Due
to its symmetry such a plasmonic crystal represents a six-port device (three
inputs and three outputs) that has been called ``tritter'' before
\cite{Zeilinger:1997}. We realized a tritter working at $\lambda_{SPP}=730$
nm, i.~e., at 750 nm laser wavelength by placing gold particles with 200 nm
diameter and 70 nm height within an area enclosed by an equilateral triangle
with $15$ $\mu$m side length. The LRM image in Fig.~3c shows that a SPP beam
(incident from the lower left) is indeed split into three reflected beams. The
identical results when using the two other inputs are shown in
Figs.~3e,g. Again, the directions of the vectors $\mathbf{f}_{i}$ can be taken
directly from the Fourier space images in Figs.~3d,f,h which show clearly the
four spots corresponding to the incident and the three reflected SPPs. The
dashed arrows correspond to a momentum transfer of $\pm\mathbf{f}_{1}$,
$\pm\mathbf{f}_{1}$, or $\pm(\mathbf{f}_{1}-\mathbf{f}_{1})$, depending on the
incident beam direction and the output beams considered (compare
Figs.~3d,f,h). We find that for the chosen geometry the plasmonic crystal
tritter distributes 85\% of the incident SPP intensity equally between the
three outputs. The remaining $\simeq 15$\% are again attributed to scattering
loss.

While all results discussed here were achieved within the visible
spectral range, plasmonic crystal devices are expected to perform
even better (e.g., in terms of spectral selectivity) in the
infrared (telecom) spectral range due to significantly lower ohmic
losses \cite{Nikolajsen:2003}. In general, the use of
multiplexers, splitters and tritters in photonic applications
might be specifically appealing due to their small footprint in
the range of $10\times10$ $\mu$m$^2$. Furthermore, the use as
building blocks for classical \cite{Teich} or quantum
\cite{Knill:2001} optical computing can be envisaged.

We acknowledge financial support from the EU under projects FP6
NMP4-CT-2003-505699 and FP6 2002-IST-1-507879.

\newpage
\begin{figure}[hbpt]
\includegraphics[width=8cm]{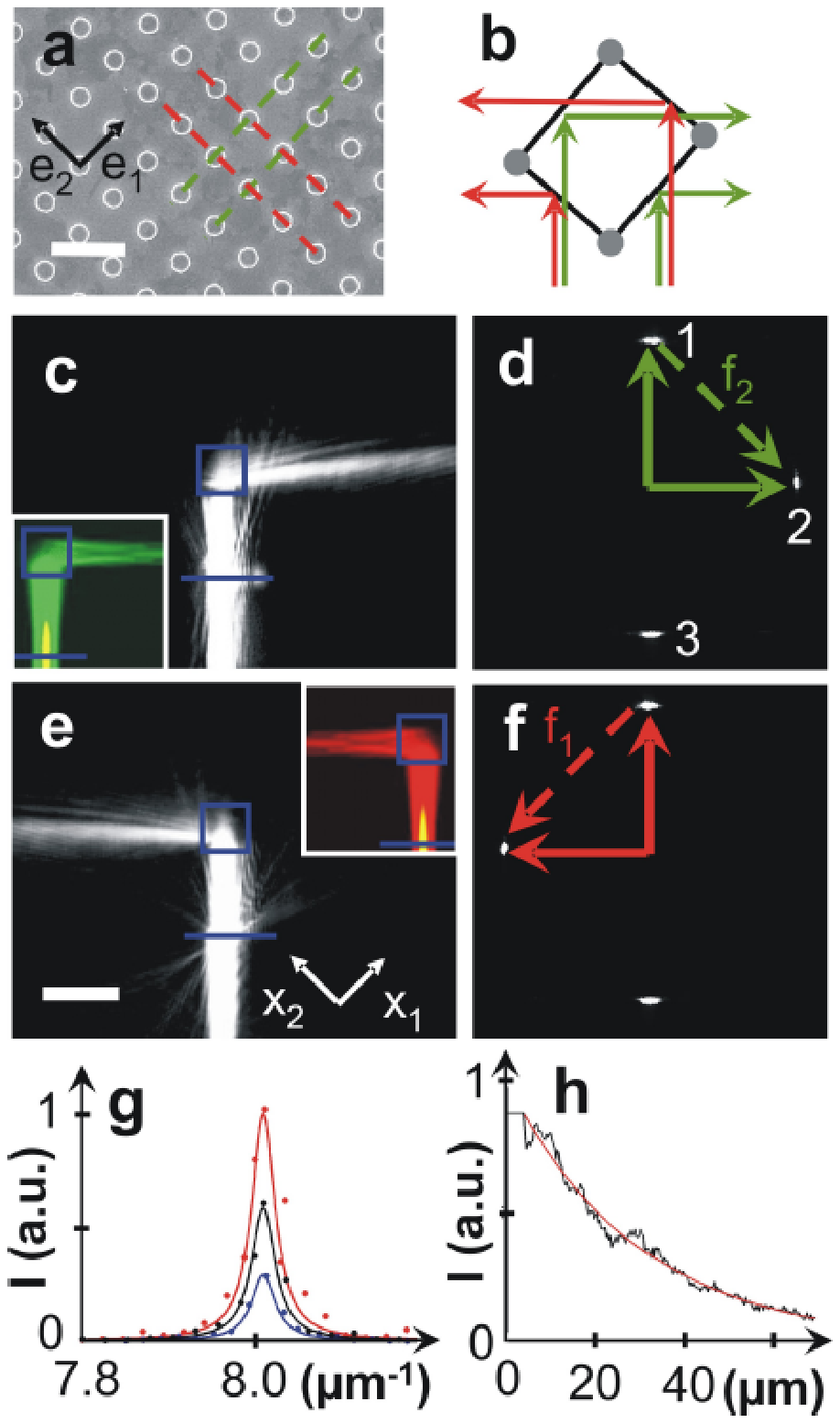}
\caption{Plasmonic crystal demultiplexer. (\textbf{a}) SEM image
of the crystal lattice built using electron beam lithography and
made of gold protrusions (200 nm diameter, 50 nm height) deposited
on a 50 nm thick gold film. ($\mathbf{e}_{1},\mathbf{e}_{2}$) is
the primitive lattice basis of the crystal and the dashed lines
show the two sets of Bragg lines corresponding to SPP wavelengths
$\lambda_{1}=784$ nm (red) and $\lambda_{2}=730$ nm (green).
(\textbf{b}) Sketch indicating SPP Bragg reflections for both
wavelengths. (\textbf{c,e}) Direct space LRM images for
wavelengths $\lambda_{2}$ and $\lambda_{1}$, respectively. The
launching ridge and the plasmonic crystal are indicated by the
blue line and the rectangle, respectively. The insets show the
according simulations. (\textbf{d,f}) Fourier space images
corresponding to \textbf{c,e}, respectively. The features 1,2,3
are discussed in the text, the continuous arrows indicate the
momenta of incident and reflected SPPs and the dashed arrow
represents the momentum transfer $\mathbf{f}_{i}$ (i=1,2) from the
Bragg mirror. (\textbf{g}) Radial cross--cuts from \textbf{d}
through 1, 2, 3 (cross--cuts are done along the short red lines in
\textbf{d}). The red, black and blue curves corresponds to
Lorentzian fits for 3,1 and 2 respectively (data points are shown
with the same color). (\textbf{h}) Angular cross--cuts from
\textbf{d} through 1, 2, 3 along the circumference of the ring
with radius $|k'_{SPP}|$ ($k'_{SPP}$ is the real part of the
wavevector associated with SPP propagating on a flat
interface)\cite{Drezet:2006}. Colors correspond to \textbf{g}.}
\end{figure}

\newpage
\begin{figure}[hbtp]
\includegraphics[width=8cm]{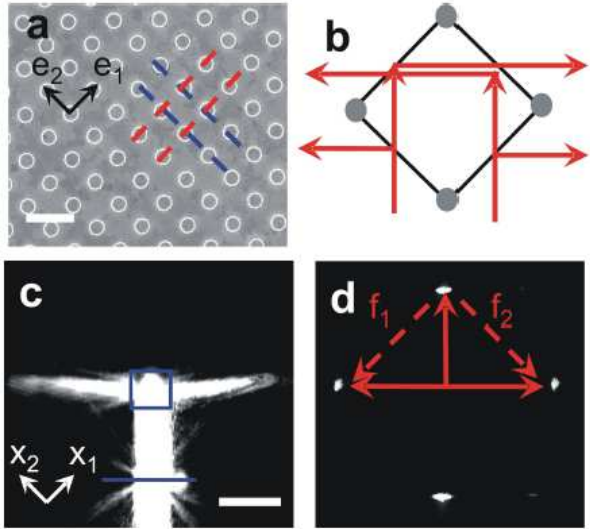}
\caption{Plasmonic crystal splitter. (\textbf{a}) SEM image,
($\mathbf{e}_{1},\mathbf{e}_{2}$) is the primitive lattice basis of the
crystal and the dashed lines show the two sets of Bragg lines corresponding to
a SPP wavelength $\lambda_{2}=730$ nm. (\textbf{b}) Sketch indicating SPP
Bragg reflection directions. (\textbf{c}) Direct space LRM image. The
launching ridge and the plasmonic crystal are indicated by the blue line and
the rectangle, respectively. (\textbf{d}) Fourier space image corresponding to
\textbf{c}. The continuous arrows indicate the momenta of incident and
reflected SPPs and the dashed arrows represent the momentum transfer
$\mathbf{f}_{i}$ (i=1,2) from the Bragg mirror.}
\end{figure}

\newpage
\begin{figure}[hbtp]
\includegraphics[width=8cm]{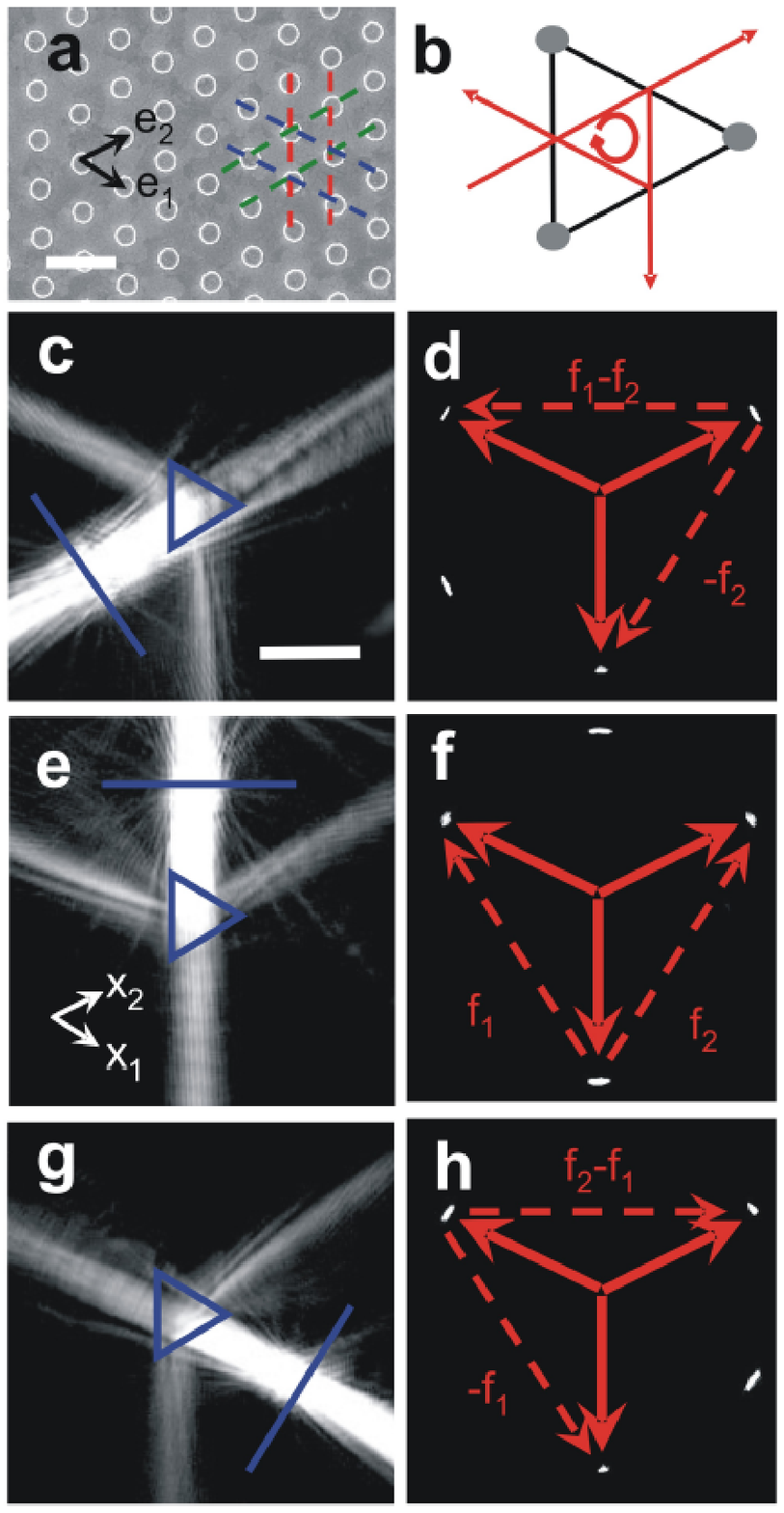}
\caption{Plasmonic crystal tritter. (\textbf{a}) SEM image,
($\mathbf{e}_{1},\mathbf{e}_{2}$) is the primitive lattice basis of the
crystal and the dashed lines show the three sets of Bragg lines corresponding
to a SPP wavelength $\lambda_{2}=730$ nm. (\textbf{a}) Sketch indicating SPP
Bragg reflection directions. (\textbf{c,e,g}) Direct space LRM images. The
launching ridge and the plasmonic crystal are indicated by the blue line and
the triangle, respectively. \textbf{d,f,h}) Fourier space images corrsponding
to \textbf{c,e,g}, respectively. The continuous arrows indicate the momenta of
the incident and reflected SPPs while the dashed arrows represent the momentum
transfer from the Bragg mirror.}
\end{figure}

\end{document}